\documentclass[twocolumn, twoside,prl]{revtex4}

\usepackage{amsmath}
\usepackage{amssymb}
\usepackage{latexsym}
\usepackage{revsymb}
\usepackage{epsfig}

\newcommand{\barr}{\begin{eqnarray}}
\newcommand{\earr}{\end{eqnarray}}

\newcommand{\beq}{\begin{equation}}
\newcommand{\eeq}{\end{equation}}

\begin{document}

\title{Quantum states and knowledge: Between pure states and density matrices.}

\author{S. Popescu$^{1,2}$}
\affiliation{$^1$ School of Physics, University of Bristol, %
 Tyndall Avenue, Bristol BS8 1TL, U.K.}

\affiliation{$^2$ Institute for Theoretical Studies, ETH Zurich}

\begin{abstract}
In this note I will present a subtle interplay between density matrices and the knowledge about their preparation, and I will argue that there is a need to consider a new type of quantum state, in between pure states and density matrices.
\end{abstract}

\date{}

\maketitle

\bigskip
\noindent
Quantum mechanics presents many crucial differences from the classical world that we know in our daily life. One of the most dramatic of them is, arguably, the one related to density matrices: In quantum mechanics it is possible to have completely different physical  situations which nevertheless cannot be distinguished from one another by any observation. This is the case of various ensembles of systems prepared in ways corresponding to the same density matrix. When information is also given about how these different situations were prepared, we can tell them apart, but as long as we only have the systems themselves, no observation can differentiate them.

\bigskip
\noindent
In this short note I will present a new and subtle interplay between density matrices and knowledge about their preparation. 

\bigskip
\noindent
Suppose Alice prepares a large ensemble of spin $1/2$ particles by taking each particle separately, tossing an unbiassed coin, and then preparing the particle either ``up'' or ``down'' along the $z$-axis. The particles are numbered, and Alice notes their preparation in a notebook;  she therefore knows the state (``up-$z$'' or ``down-$z$'') for each particle.  

\bigskip
\noindent
This ensemble of particles corresponds to the identity density matrix. 

\bigskip
\noindent
Suppose now Alice were to give this ensemble to Bob, without telling him anything about the preparation. As it is well-known, Bob has absolutely no way to distinguish this ensemble from any other ensemble that corresponds to the same density matrix (identity density matrix in this case), such as an ensemble of particles prepared by taking each particle separately, tossing an unbiased coin, and then preparing the particle either ``up'' or ``down'' along the x axis. All he can do is to is to certify that the ensemble corresponds to the identity density matrix, but he can learn nothing about which particular ensemble corresponding to that density matrix was prepared. The ``$x$-ensemble'' is physically different from the ``$z$-ensemble'' yet Bob has no way to distinguish them.

\bigskip
\noindent
Alice however, knowing what she did, can do much more than what Bob can: she can predict what results she will get if she were to measure again the particles along the $z$-axis. So, in particular, she could check if her ensemble was substituted by another ensemble corresponding to the identity density matrix. 

\bigskip
\noindent
All of the above is well known.

\bigskip
\noindent
Suppose now that Alice loses her notebook. She still remembers that she prepared the particles polarised along the z axis, but not in which direction, ``up-$z$'' or ``down-$z$'' each particle is. Apparently now she is in no better position than Bob. Indeed, exactly like Bob, she can no longer predict the outcomes of the measurements  along any axis, including the $z$-axis. Moreover, exactly like Bob, she no longer has any way to determine whether or not her ensemble was substituted by another ensemble corresponding to the identity density matrix. 

\bigskip
\noindent
As she no longer can distinguish the ensemble she created from any other ensemble corresponding to the same density matrix, we could be tempted to conclude that now, for all practical purposes, all she has is a density matrix, not a specific ensemble, exactly as Bob.

\bigskip
\noindent
Yet, Alice still has some information about the original ensemble, namely that it is a ``z-ensemble'', i.e. that each particle is polarised either ``up-z'' or ``down-z''. That represents a certain knowledge about the physical made-up of the system, and it is more than what Bob knows. But it seems that makes no difference.

\bigskip
\noindent
Surprisingly however, as I will show now, there is a difference.

\bigskip
\noindent
Consider the following scenario. After Alice prepared the ensemble, as described above, she put it in a secure place where nobody, including her, had access to it for a given time period; the particles are well isolated and remain undisturbed during this entire time. Meanwhile she ``loses'' her notebook. Actually the notebook is not lost but stolen by Charles. Charles then goes to a judge and claims that he is the one that prepared the ensemble, and that Alice had nothing to do with it.  Hence the ensemble, when released from its secure place, should be given to him. 

\bigskip
\noindent
Charles feels certain to win since, using the stolen notebook, he could prove to the judge that he knows what the ensemble is by correctly predicting the outcomes of $z$-spin measurements. He also knows that Alice has no way that she could prove to the judge that she knows what the ensemble is, since she cannot predict correctly the outcomes of any spin measurements on this ensemble.  Alice, however, can {\it disprove} Charles' claim that she has no knowledge whatsoever what the ensemble is:

\bigskip
\noindent
Hearing Charles' claim that she doesn't know what the ensemble is, Alice asks the judge not to reveal to her anything about what ensemble Charles claims to have prepared. She then tells the judge that this is a $z$-ensemble. Then she asks the judge to request Charles to disclose what the ensemble is, and to prove it by predicting the results of the spin measurements. The judge should then perform the measurements and check Charles' predictions. Of course, Charles has no option but to reveal that this is a $z$-ensemble, since if he would say anything else, he could not predict the result of the measurements. Now, the probability that Alice, just by chance, indicated the  correct spin polarisation is vanishingly small, so, by the above procedure, she clearly proves she knows what the ensemble is.

\bigskip
\noindent
An even more sophisticated scenario is also possible. Suppose Charles convinces the judge that the polarisation direction he prepared is an important commercial secret, and cannot be disclosed. Alice can still win. She asks for the spins to be measured along a direction of her choosing, and agrees to lose if Charles is able to predict the results with probability only slightly better than 1/2. 

\bigskip
\noindent
The judge agrees to this procedure since (a) Alice can only succeed if she is able to choose a direction perpendicular to the direction of polarisation; if she doesn't know the polarisation, she has vanishingly small probability of doing this, and (b) if Alice doesn't know the direction and chooses a direction at random, Charles can win without disclosing too much about the actual polarisation direction (even if he can predict the results with greater probability, he only needs to do it slightly better than 1/2, which gives him plenty of room to disguise the actual polarisation direction). 

\bigskip
\noindent
Obviously, since Alice knows that the spins were prepared polarised ``up'' or ``down'' along the $z$-axis, she requires Charles to measure along, say, the $x$-axis, where he cannot guess the results better than random and she wins.

\bigskip
\noindent
To conclude, Alices knowledge is meaningful: from her point of view the ensemble is more than the density matrix to which it corresponds, although, by herself, she cannot make any predictions better that Bob, who has no knowledge at all, and for whom the state is just a density matrix. This shows the existence of a new type of physical ``state'', which sits somewhere in between the pure state that describes the situation for Charles, who has full knowledge of the preparation, and the density matrix that describes the situation for Bob who has no knowledge at all. It is a new mathematical object that, I believe, has to be introduced in quantum mechanics.

\bigskip
\noindent
The above examples are, of course, very artificial and with no immediate application; they are however most probably just the tip of an iceberg, and I feel they have deep implications on our understanding of what quantum states are, what is physical about them, and what role our knowledge plays.

\bigskip
\noindent
{\bf Acknowledgements:} I thank Tony Short for simplifying my original example, as well as discussions with Ralph Silva, Emmanuel Zambrini and Paul Skrzypczyk. I also thank the Institute for Theoretical Studies, ETH Zurich for its support during this research.

\end{document}